\documentclass[aps,prb,twocolumn,superscriptaddress,groupedaddress,longbibliography]{revtex4-1}
\usepackage{epsfig,amsopn}
\usepackage{graphicx}
\usepackage{color}
\usepackage{amsmath,amssymb}
\usepackage{enumerate}
\newcommand\bea{\begin{eqnarray}}
\newcommand\eea{\end{eqnarray}}
\newcommand\beq{\begin{equation}}
\newcommand\eeq{\end{equation}}

\def\nn{\nonumber}
\def\f{\frac}

\def\om{\omega}

\def\si{\sigma}

\def\De{\Delta}
\def\dg{\dagger}

\def\ua{\uparrow}
\def\da{\downarrow}

\def\th{\theta}

\begin{document}
\title{Nonadiabatic charge pumping  across two superconductors 
connected through a normal metal region by periodically driven potentials} 
\author{ Abhiram Soori}  
\email{abhirams@uohyd.ac.in}
\affiliation{ School of Physics, University of Hyderabad, C. R. Rao Road, 
Gachibowli, Hyderabad-500046, India.}
\author{ M. Sivakumar}
\email{siva@uohyd.ac.in}
\affiliation{ School of Physics, University of Hyderabad, C. R. Rao Road,
Gachibowli, Hyderabad-500046, India.}
\begin{abstract} 
Periodically driven systems exhibit resonance when the difference between an
excited state energy and the ground state energy is an integer multiple of 
$\hbar$ times the driving frequency. On the other hand, when a superconducting 
phase difference is maintained between two superconductors, subgap states 
appear which carry a Josephson current. A driven Josephson junction therefore
opens up an interesting avenue where the excitations due to applied driving 
affect the current flowing from one superconductor to the other. Motivated by
this, we study charge transport in a superconductor-normal metal-superconductor~(SNS)
junction where oscillating potentials are applied to the normal metal region. 
We find that for small amplitudes of the oscillating potential, driving at 
one site reverses the direction of current at the superconducting 
phase differences when difference between the subgap  eigenenergies of the
undriven Hamiltonian is integer multiple of $\hbar$ times the driving frequency. 
For larger amplitudes of oscillating potential, driving at one site exhibits
richer features. We show that even when the two superconductors are maintained 
at same superconducting phase, a current can be driven by applying 
oscillating potentials to two sites in the normal metal differing by a phase. 
We find that when there is a nonzero Josephson current in the 
undriven system, the local peaks and valleys in current of the system driven 
with an amplitude of oscillating potential smaller than the superconducting gap 
indicates sharp excitations in the system. 
In the adiabatic limit, we find that charge transferred in one time period diverges 
as a powerlaw with pumping frequency when a Josephson current flows in the 
undriven system. Our calculations are  exact and can 
be applied to finite systems. We discuss possible experimental setups where 
our predictions can be tested. 
\end{abstract}
\maketitle

\section{Introduction}
DC Josephson effect is a phenomenon of flow of current across two bulk superconductors 
maintained at a phase difference~\cite{josephson62}. This current is directly
proportional to the sine of the 
phase difference. A simple model to study this phenomenon consists of  two semi-infinite
one-dimensional superconducting channels connected by a suitable boundary condition that 
describes the junction. In such a model where the superconducting channels are described
by Bogoliubov-de-Genne mean-field Hamiltonian, the bulk states in the continuum band do
not contribute to the Josephson current and the entire Josephson current is carried by
the lower (at energy $-E_b$) of the two quasiparticle bound states formed (at 
energies $\pm E_b$)  within the superconducting gap~\cite{furusaki99}. 

The Josephson effect is important from the point of view of fundamental physics
as well as applications. Few years after its experimental confirmation in 1963~\cite{anderson63},
the value of $2e/h$ was experimentally measured~\cite{parker67}. The prediction by Josephson
that a weaklink between two superconductor irradiated with  microwave radiation can convert
the frequency of the radiation into  dc voltage was verified by Shapiro~\cite{shapiro63} 
and this eventually led to the development of voltage standard based on Josephson 
effect~\cite{levinsen77}. A Josephson junction embedded in a low temperature surrounding 
can sense the thermal noise
through frequency modulations and this principle has been used to measure temperatures of
the order of micro-milli Kelvin~\cite{Kamper71}.  Nontrivial spin 
triplet superconductivity has been engineered and observed in Josephson junctions 
made of ferromagnetic material~\cite{khaire10}. Also Josephson junctions with novel 
materials such as graphene~\cite{Heersche2007} and topological insulators~\cite{Williams12}
have been studied experimentally. 
Recently, the existence of Majorana fermions in topological semiconductor 
quantum wires was confirmed by their unusual $4\pi$-Josephson effect in contrast to the
$2\pi$-Josephson effect exhibited by the nontopological superconductors~\cite{rokhi2012,Laroche2019}.
Josephson junctions emit radiation when a current above a critical current is passed
through them and this feature is used in developing Terahertz frequency radiation sources
from them~\cite{Kleiner07}. Quantum bits have been created using Josephson junctions 
and these are building blocks of quantum computer~\cite{wendin07,Clarke2008}.

 Quantum charge pumping is a nonequilibrium phenomenon where charge
 is transferred from one reservoir to the other by the application of time dependent 
 potentials in the transport channel~\cite{thouless83,brouwer98,switkes99,brouwer01,
 avron01,moskalets02,agarwal07,agarwal07prb,soori10}. While charge pumped 
 adiabatically in one time period
 is quantized~\cite{thouless83}, an interesting extension to fractional quantization 
 of charge in a fraction of time period in certain models has also been 
 shown~\cite{marra15}. 
 The phenomenon of pumping purely spin in a mesoscopic device 
 is also explored~\cite{citro06}. Also, pumping in novel materials such as silicene
 has been studied recently~\cite{paul17}.
 Recently, the idea of pumping has been invoked in 
 distinguishing Majorana bound states from Andreev bound states~\cite{tripathi19}.
 Following the work on pumping between a normal metal lead and a superconducting
 lead~\cite{blaauboer02}, adiabatic pumping in 
 superconductor-normal~metal-superconductor~(SNS)
 weaklink was studied~\cite{governale05}. Quantum  charge pumping is a special case
 of Floquet dynamics -a study of quantum systems with periodic driving which has 
 seen a  lot of activity in the last decade~\cite{grifoni98,russomanno12,laza14prl,
 laza14,moessner17,oka19,giovannini19}. It was shown that periodically driven
 isolated many body systems approach periodic steady 
 state after an initial transient dynamics~\cite{laza14prl,russomanno12}. 
 In this work, we study nonadiabatic pumping in an SNS structure when oscillating potentials
 are applied to the normal~metal region. 
 
 In the adiabatic limit, two parameters in the system need to be periodically varied
 to pump charge~\cite{brouwer98}. However, in the nonadiabatic limit it was shown that
 by breaking spatial inversion symmetry, one can pump charge by varying only one 
 parameter periodically in the system~\cite{soori10}. An interesting phenomenon in this
 context is assisted pumping. Resonant levels are created in a region by static potentials
 and an oscillating potential is applied to the left of this region. As the Fermi energy
 is varied, pumped current shows a peak when the  difference between a resonant energy and 
 the Fermi energy is equal to $\hbar$ times the frequency of the oscillating potential. 
 This is because  interaction of an electron at the Fermi energy with the oscillating
 potential results  in change in the  electron's energy by $\hbar \omega$ as depicted
 in ref.~\cite{agarwal07prb}. This motivates us to study current in a Josephson junction
 with an oscillating potential in between the two superconductors. A simple question to
 ask would be- ` what is the effect of oscillating potential on the Josephson current,
 particularly when the difference between the two subgap bound state energies is equal
 to $\hbar$ times the pumping frequency $\omega$?' We show that for small amplitudes
 of oscillating 
 potential, when the ratio $2E_b/\hbar \omega$ is a positive integer, the current deviates
 from  Josephson current of the undriven system. For larger 
 amplitudes of the oscillating potential, we find that the above said deviation of the
 current  happens even when $2E_b/\hbar \omega$ is not an integer. We explain these 
 deviations  in the current. 
 
 A superconductor that is grounded acts as a reservoir for charge. A system described by
 lattice model with finite number of sites can be exactly studied by exact diagonalization
 of the Hamiltonian. Further, when an oscillating potential is applied the system 
 is disturbed from 
 the equilibrium ground state and the current is not periodic though the Hamiltonian is. 
 Hence, current averaged over one time period is not a measure of pumped charge. Instead, 
 current averaged over infinite time starting from the moment oscillating potential is 
 switched on is a measure of pumped charge. We call this current `time averaged current'
 and denote it by $I_{av}$. It was shown that by suitably choosing the basis, this 
  current averaged over infinite time can be reduced to current averaged over one time 
 period~\cite{soori10}. This is a numerical calculation and requires the Hilbert 
 space dimension to be finite.  Hence we choose the superconductors to have finite 
 number of lattice sites. 
 
 Rest of the paper is structured as follows. Sec.~\ref{sec-model} discusses the model.
 Sec.~\ref{sec-calc} discusses the details of calculation.
 Sec.~\ref{sec-res} discusses the results and presents an analysis of the results. 
 Finally we discuss the implications of our results, experimental setups where our results 
 can be put to test and conclude in sec.~\ref{sec-con}.
 \section{Model}~\label{sec-model}
\begin{figure}
 \includegraphics[width=8cm]{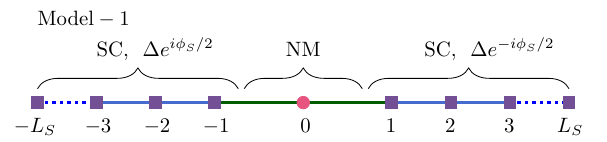}
 \includegraphics[width=8cm]{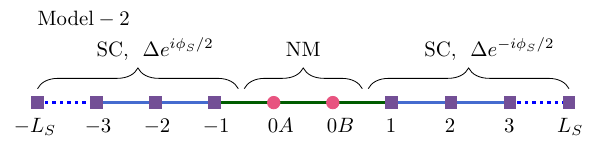}
 \caption{Schematic diagram of the two models studied in this work. In model-1, the normal
 metal~(NM) consists of only one site denoted by $0$ where time-dependent potential $V(t)$
 is applied. In model-2, NM consists of two sites denoted by $0A$ and $0B$ where the time 
 dependent potentials $V_1(t)$ and $V_2(t)$ are applied respectively. In both the models, 
 two finite superconductors are attached to either sides of NM differing by a superconducting
 phase $\phi_S$.}~\label{fig-schem}
\end{figure}

 Each superconductor in the SNS junction  consists of $L_S$ sites. The superconductor 
 on the left has a phase $\phi_S/2$ and the 
 superconductor on the right has a phase $-\phi_S/2$. We take two models for the 
 normal metal in between. In the first model, the normal metal in between  consists 
 of only one site and an oscillating potential 
 $V(t)=V_0\cos{(\om t +\phi_{0V})}\cdot\Theta(t)$  is applied on it. In the second model,
 the normal metal in between has two sites. An oscillating potential 
 $V_1(t)=V_0\cos{(\om t +\phi_{0V})}\cdot\Theta(t)$ is applied to the first site and 
 an oscillating potential $V_2(t)=V_0\cos{(\om t +\phi_{0V}+\delta\phi_{V})}\cdot\Theta(t)$
 differing  by a phase factor of $\delta\phi_V$ is applied to the second site. 
 In both the models, the oscillating potential is zero until $t=0$ and for $t>0$, 
 it is sinusoidal with a frequency $\om$. A static chemical potential $\mu$ is 
 present on both the  superconductors and on the normal metal sites. The 
 Hamiltonian for the system in the first model is given by $H=H_0+H_1(t)$, where
 \bea  
 H_0 &=& H_L + H_{LN} + H_N + H_{NR} + H_R, \nn \\
 H_L &=& -w\sum_{n=-1}^{-L_S+1}(c^{\dg}_{n-1}\tau_zc_{n}+{\rm h.c.})
 +\sum_{n=-1}^{-L_S}c^{\dg}_{n}\big[-\mu \tau_z\nn\\
 &&+\De\cos{(\f{\phi_S}{2})}\tau_x+\De\sin{(\f{\phi_S}{2})}\tau_y \big]c_{n}, \nn \\
H_R&=&-w\sum_{n=1}^{L_S-1}(c^{\dg}_{n+1}\tau_zc_{n}+{\rm h.c.}) 
+\sum_{n=1}^{L_S}c^{\dg}_{n}\big[-\mu \tau_z\nn\\ 
&&+\De\cos{(\f{\phi_S}{2})}\tau_x-\De\sin{(\f{\phi_S}{2})}\tau_y \big]c_{n}, \nn \\
H_N&=& -\mu c^{\dg}_{0}\tau_zc_{0}, \nn\\
H_{LN}&=&-w'(c^{\dg}_{-1}\tau_zc_{0}+{\rm h.c.}),\nn\\
H_{NR}&=&-w'(c^{\dg}_{1}\tau_zc_{0}+{\rm h.c.}),\nn\\
H_1(t)&=&V(t) c^{\dg}_{0}\tau_zc_{0},~~\label{eq:H-1site}
 \eea
 $c_n=[c_{n,\ua},-c_{n,\da},c^{\dg}_{n,\da},c^{\dg}_{n,\ua}]^T$, where $c_{n,\si}$ 
 annihilates an electron at site $n$ with spin $\si$. $\tau_{x,y,z}$ are the 
 Pauli matrices acting in the particle-hole sector. The Hamiltonian for the second model
 which describes the phenomenon of pumping at two sites is given by $H=H_0+H_1(t)$
 resembling closely with eq.~\eqref{eq:H-1site} except for the following changes
 \bea 
 H_N&=& -\mu (c^{\dg}_{0A}\tau_zc_{0A}+c^{\dg}_{0B}\tau_zc_{0B})
            -w'' (c^{\dg}_{0A}\tau_zc_{0B}+{\rm h.c.})  , \nn\\
H_{LN}&=&-w'(c^{\dg}_{-1}\tau_zc_{0A}+{\rm h.c.}),\nn\\
H_{NR}&=&-w'(c^{\dg}_{1}\tau_zc_{0B}+{\rm h.c.}),\nn\\
H_1(t)&=&V_1(t) c^{\dg}_{0A}\tau_zc_{0A}+V_2(t) c^{\dg}_{0B}\tau_zc_{0B}.~\label{eq:H-2site}
 \eea
A schematic picture of the two models is shown in Fig.~\ref{fig-schem}. 

 \section{Details of calculation}~\label{sec-calc}
 To begin with, just a finite phase difference $\phi_S$ in absence of oscillating 
 potential drives a Josephson current $I_J$ from one superconductor to the other. 
 This Josephson current can be calculated by summing over expectation values of 
 current operator at the bond between superconductor and the normal metal 
 [$(-1,0)$ for the first model and $(-1,0A)$ for the second model]
 taken with respect to all eigenstates of
 $H_0$ which have energies less than zero. In the limit of large system size
 ($L_S\to\infty$), only the (subgap) bound state with energy $-E_b$ less than
 zero contributes to the Josephson current. For a finite system, though the subgap 
 state with energy $-E_b$ contributes the significantly to the Josephson current, 
 the contribution from other occupied states below the Fermi energy cannot be 
 neglected.  Let $\{|u_i\rangle,E_i,~i=1,2,..,N\}$ (where $N$ is the dimension
 of the Hilbert space) be the
 eigenstates and eigenenergies of $H_0$ arranged in the ascending order of the 
 eigenenergies. For the first model which has one normal metal site in the middle, 
 $N=4(2L_S+1)$ and for the second model which has two normal metal sites in the middle, 
 $N=8(L_S+1)$. For the first model, the current operator at bond $(-1,0)$ is 
 \beq \hat J = \f{-iew'}{\hbar}(c^{\dg}_{-1}c_{0}-c^{\dg}_{0}c_{-1}).
 \label{eq:curr-op1}\eeq
 For the second model, the current operator at bond $(-1,0A)$ is 
 \beq \hat J = \f{-iew'}{\hbar}(c^{\dg}_{-1}c_{0A}-c^{\dg}_{0A}c_{-1}).
 \label{eq:curr-op2}\eeq
 Here, $e$ is electron charge.
 The Josephson current of the undriven system is sum of currents carried by the $N/2$
 occupied states of the Hamiltonian $H_0$ and is given by
 \beq I_J = \sum_{i=1}^{N/2} \langle u_i|\hat J|u_i\rangle. \label{eq:jo-curr}\eeq
 
 A time dependent potential switched on at time $t=0$ takes the system 
 away from equilibrium ground state. Though the Hamiltonian is now periodic in time, 
 the current is not periodic. Hence, current averaged over one time period 
 does not quantify the charge that is transferred from one superconductor
 to the other. On the other hand, current averaged over infinite time starting 
 from $t=0$ is a good measure of charge that is transferred from one superconductor
 to the other. It was shown that this  infinite time average can be reduced to an 
 average over one time period in the following way~\cite{soori10}. Let the 
 time interval $[0,T]$ be divided into $M$ equal slices. Here, $T=2\pi/\om$ is 
 the time period of the oscillating potential. Size of each slice
 is $dt=T/M$. Let $t_k$ be at the center of $k$-th interval. Then, the unitary
 time evolution operator from $t=0$ to $t=T$ under the discretized oscillating 
 potential is
 \beq U(T,0) = {\cal T} \prod_{k=1}^M {\rm exp}[-i H(t_k)dt], \label{eq:UT} \eeq
 where ${\cal T}$ is time ordering operator that moves the operator at earlier time 
 to the right and the potential $V(t)$ is taken to be equal to $V(t_k)$ in the $k$th
 time interval. Let $|v_j\rangle$ be the eigenstate of $U(T,0)$ with eigenvalue
 $e^{i\th_j}$. Such eigenstates are called Floquet states. When $\{\th_j,~j=1,2,..,N\}$
are nondegenerate, the  time averaged current $I_{av}$ is given by 
\bea I_{av} &=& \sum_{i=1}^{N/2} \sum_{j=1}^N |c_{ij}|^2 (J_T)_{jj} \nn \\
{\rm where~~} c_{ij} &=& \langle u_i|v_j\rangle \nn \\ 
{\rm and~~}(J_T)_{jj} &=& \f{1}{T}\sum_{k=1}^M\langle v_j|U^{\dg}(t_k,
0)|\hat J|U(t_k,0)|v_j\rangle dt. \label{eq:curr} \eea
Physically, one can see from eq.~\eqref{eq:curr} that current is carried by the Floquet
states $|v_j\rangle$. The sum over $i=1,2,..,N/2$ here means that contributions from 
all the initially occupied states $|u_i\rangle$ are counted.  For later use, we shall 
define $I_{b,av}=\sum_{j=1}^N |c_{N/2,j}|^2 (J_T)_{jj}$ -the current contribution from
the initially occupied subgap eigenstate of $H_0$ with eigenenergy $-E_b$. This is going
to play a major role in our analysis as $I_{b,av}$ is substantially high in $I_{av}$. 

\section{Results and Analysis}~\label{sec-res}
\subsection{One-site pumping with small $V_0$}~\label{sec-res-small-V0}
\begin{figure}
 \includegraphics[width=8cm]{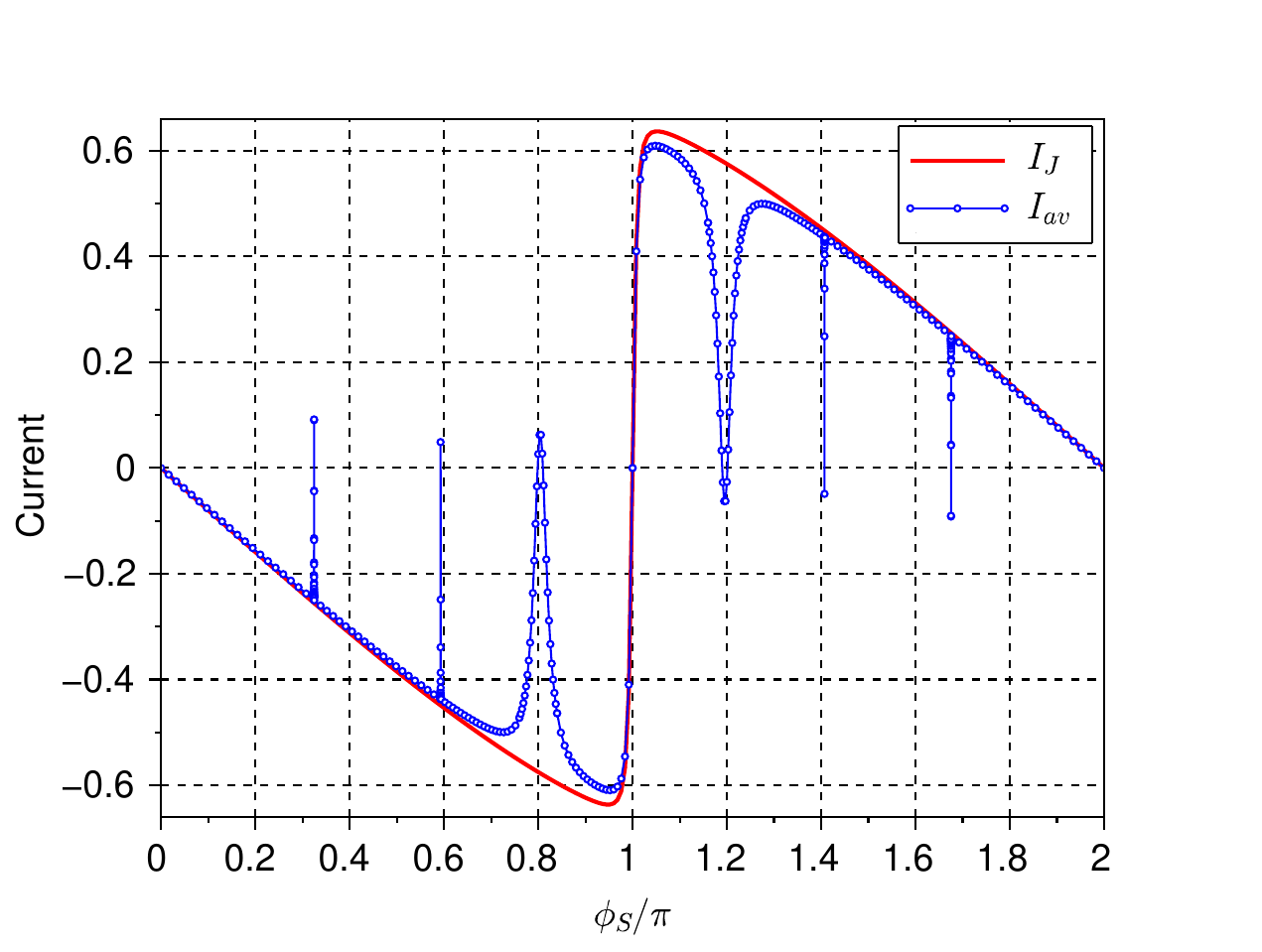}
 \caption{Current in units of $ew/\hbar$ between the two superconductors 
 plotted versus the Josephson phase
 difference. $I_J$ is the equilibrium Josephson current before the oscillating potential 
 is switched on. $I_{av}$ is the  time averaged current.
 Parameters: $L_S=4$, $\mu=0.01w$, $\De=0.5w$, $w'=0.9w$, $V_0=0.05w$, 
$\phi_{0V}=-0.5\pi$ and $\hbar \om = 0.2w$.
 }~\label{fig:curr1}
\end{figure}

We first study one site pumping - the phenomenon of charge transport when oscillating
potential is applied to only one site. We choose the amplitude of pumping potential 
$V_0$ to be small compared to the superconducting gap $\De$. 
For the system described by the Hamiltonian in eq.~\eqref{eq:H-1site}, the following 
parameters are chosen: $L_S=4$, $\mu=0.01w$, $\De=0.5w$, $w'=0.9w$, $V_0=0.05w$, 
$\phi_{0V}=-0.5\pi$ and $\hbar \om = 0.2w$. The time period $T=2\pi/\om$. The time 
interval $[0,T]$ is sliced into $M=50$ slices. $V_0$ is chosen to be small 
so that the subgap states do not mix with the bulk states of the superconductor.
$\De=0.5w$ is chosen to be substantially large so that the wavefunction of the subgap
energy state decays fast into the bulk of the superconductor so that we can work 
with a smaller superconductor. The zero energy wavefunction on last site for a system
with $L_S=4$ has a magnitude which is $1/e$ times its magnitude at the center.
In fig.~\ref{fig:curr1}, we plot 
the Josephson current and the  time averaged current as a function of 
the superconducting phase difference $\phi_S$. The currents are antisymmetric 
about the phase difference $\phi_S=\pi$. At values 
$\phi_S/\pi=0.3246,0.5929,0.8045,(2.0-0.8045),(2.0-0.5929),(2.0-0.3246)$, the 
 time averaged current deviates the most from the Josephson current. 
At these values of phase difference, $I_{av}$ is very close to zero and
$I_{b,av}$ is even more closer to zero. We explain the origin of these 
deviations in the following paragraph.

Since the subgap energy state at $-E_b$ is closest to the Fermi energy, this state 
participates the most in transport. As explained by the Floquet theory of 
pumping~\cite{moskalets02},  $H_1(t)$ mixes the state at energy $E$ with 
the states at energies $E+n\hbar\om$
where $n$ is any integer. As $\phi_S$ varies, the difference $2E_b$ between 
the lowest unoccupied state of $H_0$ (at energy $E_b$) and the highest occupied
state of $H_0$ (at energy $-E_b$)  changes. When this difference is an integer 
multiple of $\hbar\omega$, the transitions mediated by $H_1(t)$ take place 
between the energy levels $\pm E_b$ and such a transition is the reason for the
deviation of $I_{av}$ from $I_J$. 
\begin{figure}
 \includegraphics[width=8cm]{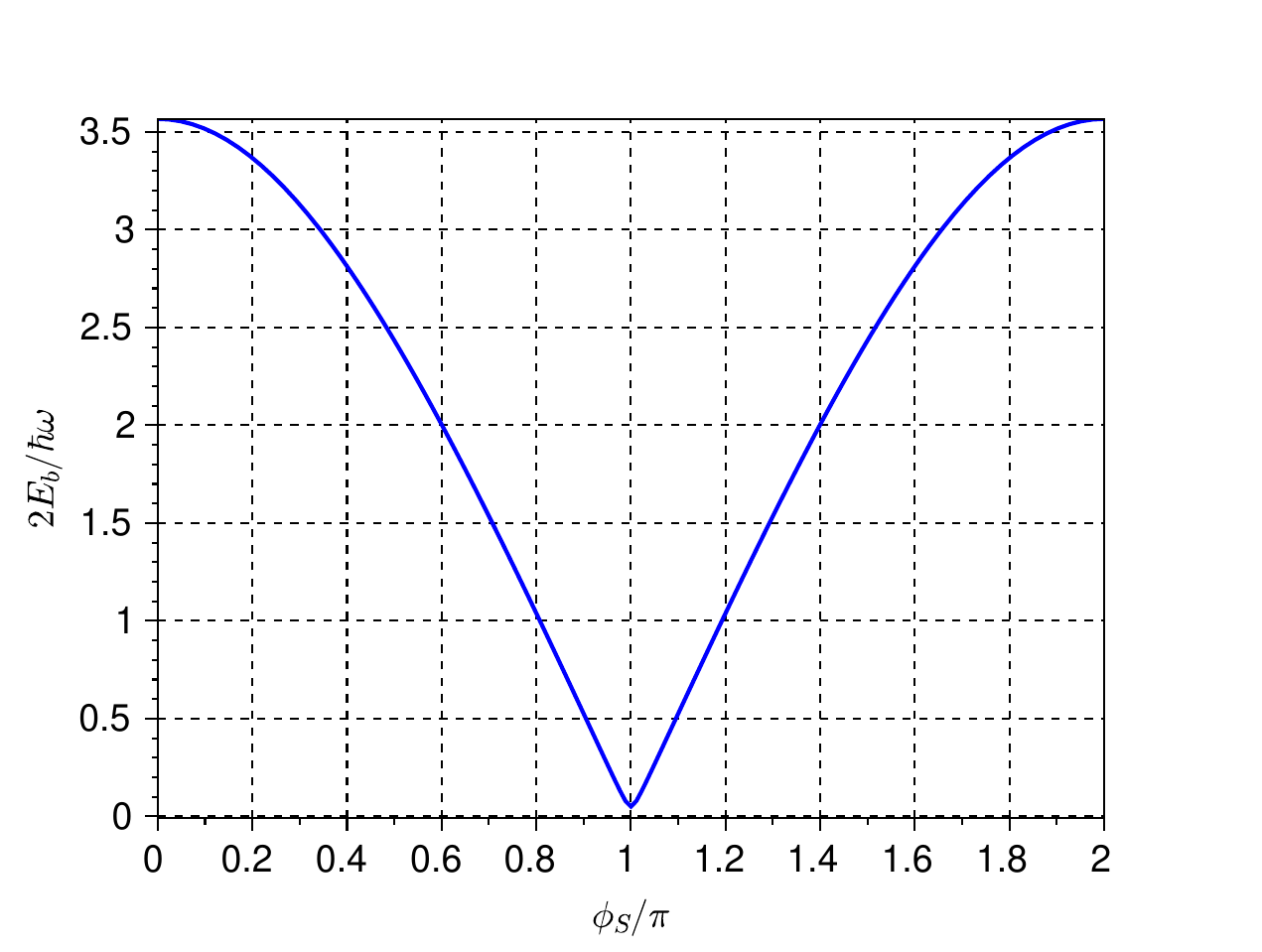}
 \caption{$2E_b/\hbar \omega$ plotted versus the Josephson phase difference, where 
 $\pm E_b$ are the subgap energies and $\omega$ is the frequency of the time
 dependent potential. }~\label{fig:e-phi}
\end{figure}
In Fig.~\ref{fig:e-phi}, we plot $2E_b/\hbar\om$ versus $\phi_S$ and find that
at the values of $\phi_S$ where the current $I_{av}$ saw substantially high
deviation from $I_J$, $2E_b/\hbar\om$ is an integer. 

These deviations in the  time averaged current $I_{av}$ from the Josephson
current $I_J$ can be understood in another way. Let us focus on the current 
contribution $I_{b,av}$ due to the subgap state  $|u_{N/2}\rangle$ . 
Since $I_{b,av}=\sum_{j=1}^N |c_{N/2,j}|^2 (J_T)_{jj}$, the current carried by 
the Floquet state $j_0$ for which the overlap $|c_{N/2,j_0}|^2$ is significant
contributes the most to $I_{b,av}$. At any value of $\phi_S$ where the deviation of
$I_{av}$ from $I_J$ is small, we find that the overlap $|c_{N/2,j_0}|^2\sim 1$. 
This means that the Floquet state $|v_{j_0}\rangle$
is very close to the subgap state $|u_{N/2}\rangle$ and hence the current it carries
is almost equal to the Josephson current. At values of $\phi_S$ where the deviation 
of $I_{av}$ from $I_J$ is significant, we find that more than one Floquet state
has significant overlap with the initial state $|u_{N/2}\rangle$. For the case of 
$\phi_S=0.8045\pi$, $|c_{N/2,1}|^2=0.50614$ and $|c_{N/2,2}|=0.49386$ meaning that there
are two Floquet states $|v_1\rangle$ and $|v_2\rangle$ which have high overlap with 
the initial subgap state. The currents carried by these Floquet states are
$(J_T)_{11}=0.004472ew/\hbar$ and $(J_T)_{22}=-0.004472ew/\hbar$ and these currents
almost cancel out while calculating $I_{b,av}$ due to the almost equal probabilities
multiplying them. Therefore, $I_{b,av}\sim 1.094\times10^{-4}ew/\hbar$ (there is a factor of two 
multiplying the current here due to spin which we did not take into account while 
counting the states). Each of the two Floquet states that matters here has significant 
and almost equal overlap with the two subgap states $|u_{N/2}\rangle$ and 
$|u_{N/2+1}\rangle$ at energies $-E_b$ and $+E_b$. Hence the eventual Floquet states
to which the ground state is driven into are  (almost)~equal superpositions of the 
ground state $|u_{N/2}\rangle$ and the excited state $|u_{N/2+1}\rangle$. Since the
ground state and excited state are at energies $\mp E_b$, they are related by 
particle-hole symmetry. Hence they carry equal and opposite currents. Hence it can be 
heuristically said that the total current carried by the eventual nonequilibrium state 
which is  an almost equal superposition of the ground state and the excited state 
is (almost)~zero. Similarly, the deviation of $I_{av}$ from $I_J$ at other values of the
Josephson phase difference can be explained. 

\subsection{One site pumping with larger $V_0$ }~\label{sec-res-V0-delta}
\begin{figure}
 \includegraphics[width=8cm]{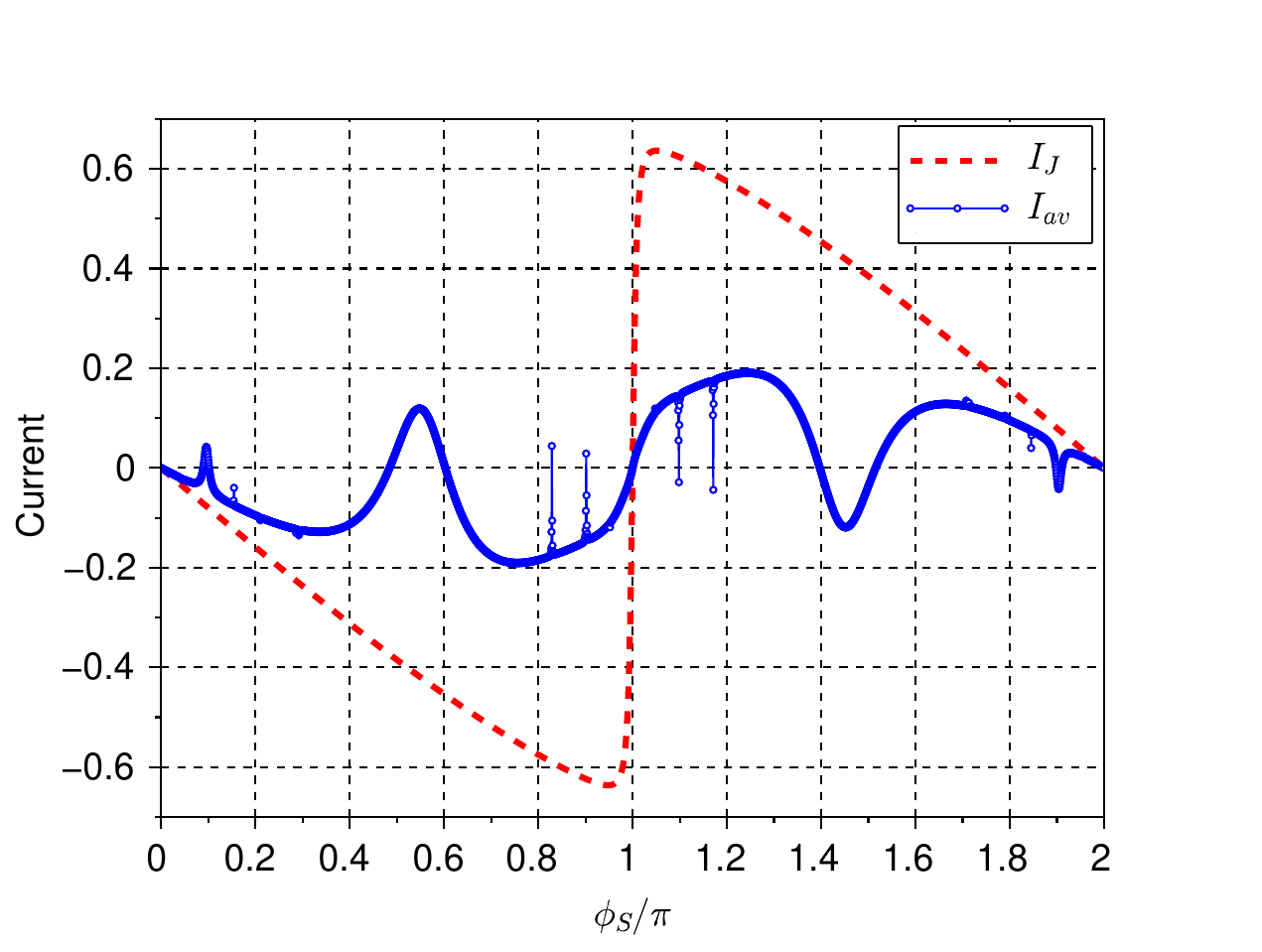} 
 \caption{Time averaged current $I_{av}$ for the driven system and the 
 Josephson current $I_J$ for the undriven system plotted as a function of the 
 superconducting phase difference $\phi_S$. Parameters chosen are same as for
 Fig.~\ref{fig:curr1} except for $V_0$ which takes the value $V_0=w$.}\label{fig:curr2}
\end{figure}

Now, we make one change to the parameters compared to the previous subsection. We 
choose $V_0=w$. This means the oscillating potential $V(t)$ helps to access 
higher energy eigenstates of $H_0$ starting from the ground state. 
Hence richer features are 
expected in the  time averaged current. In Fig.~\ref{fig:curr2} we plot the 
currents $I_{av}$ and $I_J$ as a function of the phase difference $\phi_S$. 
Substantial deviation of $I_{av}$ from $I_J$ is observed  in the entire range 
$0<\phi_S/\pi<2$ except at $\phi_S=0,\pi$. Furthermore, local peaks/valleys are observed
at $\phi_S/\pi=0.096, 0.1548, 0.2105, 0.2866, 0.2925, 0.5482, 0.8289, 0.9016,$ 
$(2.0-0.9016), (2.0-0.8289), (2.0-0.5482), (2.0-0.2925), (2.0-0.2866), (2.0-0.2105),
(2.0-0.1548), (2.0-0.096)$. 
This deviation in current is obviously due to mixing between eigenstates of $H_0$
caused by $H_1(t)$. Initially at time $t<0$, the system is in the ground state with 
all negative energy states being occupied. After $t=0$, $H_1(t)$ causes excitations. A
measure of excitations is the deviation $[-(E-E_0)/E_0]$ in the  time averaged
energy $E$ relative to the ground state energy $E_0$,  where 
\bea E &=& \sum_{i=1}^{N/2} \sum_{j=1}^N |c_{ij}|^2 (E_T)_{jj},~~~
 c_{ij}~=~ \langle u_i|v_j\rangle, \nn \\ 
(E_T)_{jj} &=& \f{1}{T}\sum_{k=1}^M\langle v_j|U^{\dg}(t_k,
0)|\hat H_0|U(t_k,0)|v_j\rangle dt, \nn \\
{\rm and ~~}E_0 &=& \sum_{i=1}^{N/2} \langle u_i|H_0|u_i\rangle.~\label{eq:ET}
\eea
A negative sign multiplies the numerator here since the denominator $E_0$ is negative.
\begin{figure}
 \includegraphics[width=8cm]{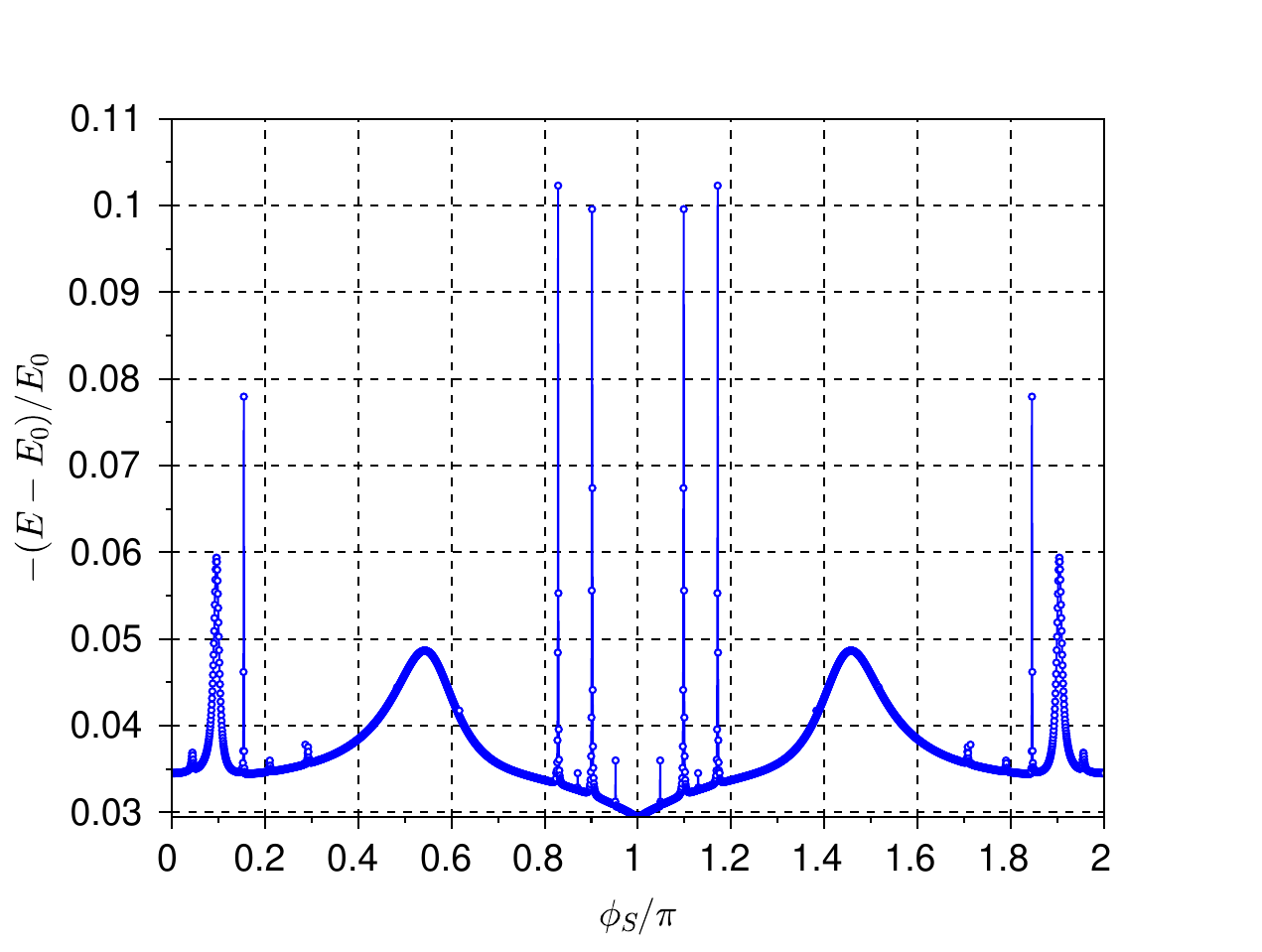}
 \caption{ Deviation in the long time averaged energy $E$ relative to the ground state
 energy $E_0$ of the time independent Hamiltonian. Parameters chosen are same as for
 Fig.~\ref{fig:curr2}.}\label{fig:dE}
\end{figure} 
This relative deviation in energy versus the superconducting phase difference $\phi_S$
is plotted in Fig.~\ref{fig:dE}. A comparison between
Fig.~\ref{fig:curr2} and Fig.~\ref{fig:dE} indicates that deviation in 
the current $I_{av}$ from $I_J$ is a fingerprint of excitations in the Floquet system. 
Thus, a deviation in current $I_{av}$ from $I_J$ can be seen as a sign of excitations
but not as a direct quantitative measure. A local peak or a local valley in $I_{av}$
is a sign of an excitation in the driven system. 

\subsection{Two site pumping with $\phi_S=0$}
\begin{figure}
 \includegraphics[width=8cm]{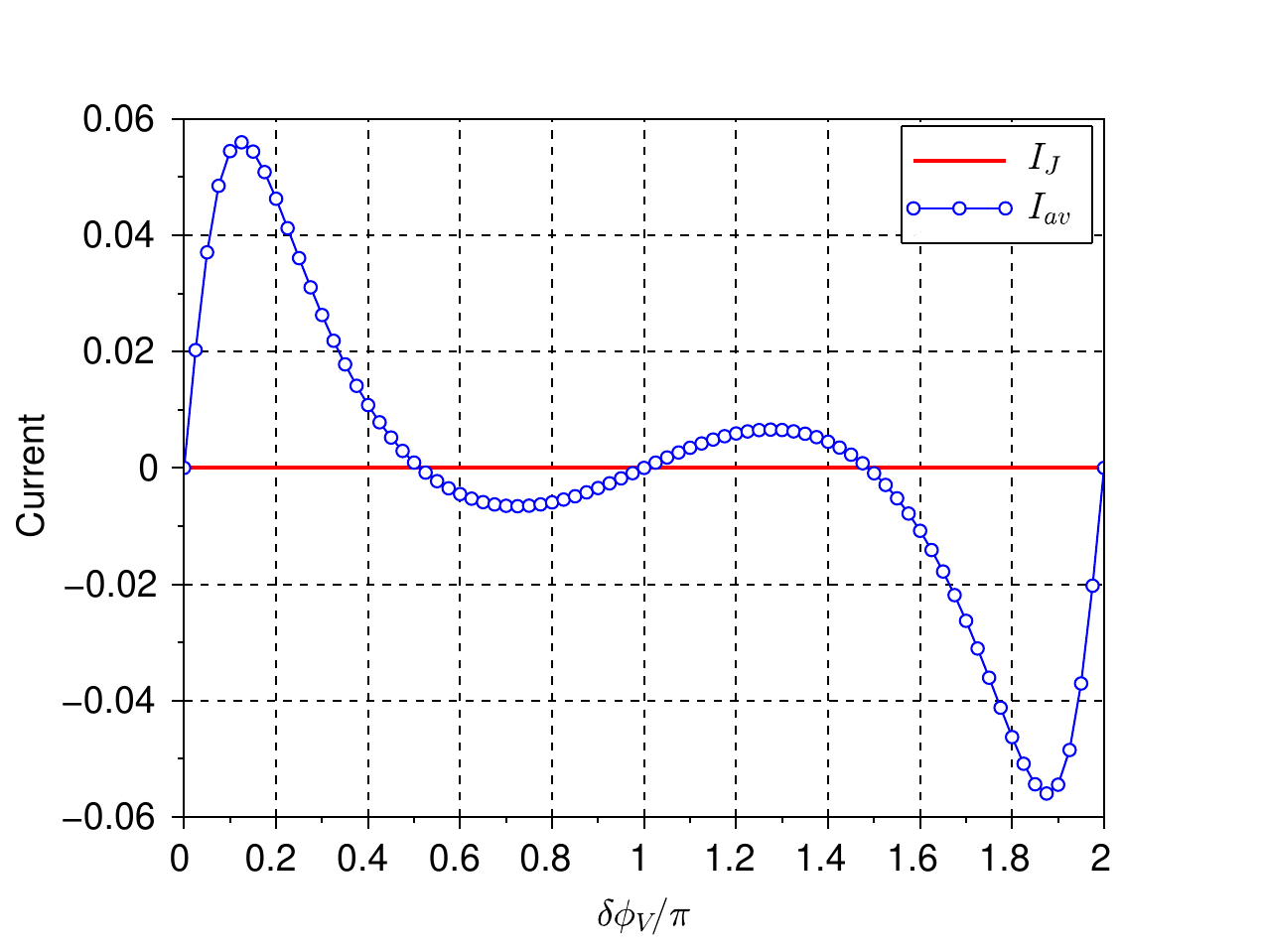}
 \caption{Time averaged current $I_{av}$ in units of $ew/\hbar$ versus the 
 phase difference $\delta\phi_V$  between the applied oscillating potentials 
 for two site pumping. 
 Parameters: 
 $L_S=4$, $\mu=0.01w$, $\De=0.5w$, $w'=0.9$, $w''=0.3w$, $V_0=0.05w$, 
 $\phi_{0V}=0$, $\phi_S=0$,  $\om=2E_b$ and the time interval of one time 
 period $[0,2\pi/\om]$ is divided into  50 intervals of equal size for
 calculating $U(T,0)$.}~\label{fig:curr3}
\end{figure}

We now turn to two site pumping -the phenomenon of charge transport when oscillating
potentials are applied to two sites in the normal metal region. For the system described by the 
Hamiltonian in eq.~\eqref{eq:H-2site}, the parameters are chosen to be 
$L_S=4$, $\mu=0.01w$, $\De=0.5w$, $w'=0.9$, $w''=0.3w$, $V_0=0.05w$, 
 $\phi_{0V}=0$, $\phi_S=0$,  $\om=2E_b$ and the time interval of one time 
 period $[0,2\pi/\om]$ is divided into  50 intervals of equal size for
 calculating $U(T,0)$. First we check whether there is charge transport purely due to 
 oscillating potentials when the superconducting phase difference between the two 
 superconductors is zero ($\phi_S=0$). The pumping frequency is set equal to the energy
 difference between the lowest unoccupied energy level of $H_0$ ($E_b$) and the 
 highest occupied energy level of $H_0$ ($-E_b$). The pumped current which is 
  time averaged current versus the phase difference between the oscillating
 potentials $\delta\phi_V$ is plotted in Fig.~\ref{fig:curr3}.
The current can be pumped in either directions depending upon the difference in phases 
of the pumping potentials. However, none of the eigenstates of $H_0$ carry any nonzero 
current since the superconducting phase difference is zero. This means that even when
there are excitations in the system, there may not be a change in the current. Hence,
the pumped current $I_{av}$ cannot taken as a measure of excitations when $\phi_S=0$.
By the same logic, a nonzero pumped current does not signify excitations when $\phi_S$
is an integer multiple of $\pi$. 
\subsection{Two site pumping with $\phi_S=0.9\pi$}
\begin{figure}
 \includegraphics[width=8cm]{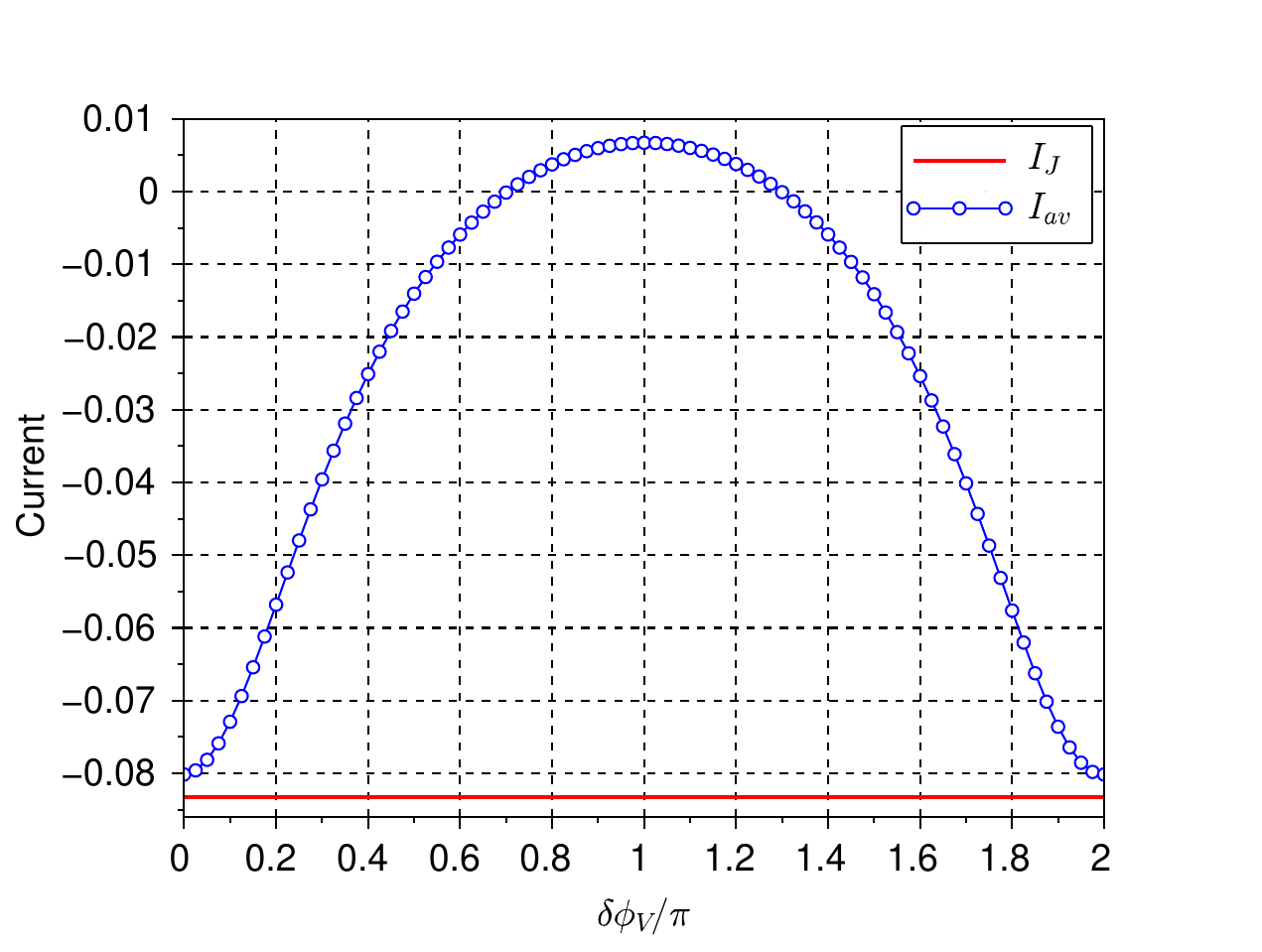}
 \caption{Time averaged current $I_{av}$ in units of $ew/\hbar$
 versus the phase difference $\delta\phi_V$
 between the applied oscillating potentials for two site pumping.
 Josephson current $I_J$ of the undriven system is plotted for reference. 
 Parameters: 
 $L_S=4$, $\mu=0.01w$, $\De=0.5w$, $w'=0.9w$, $w''=0.3w$, $V_0=0.05w$, 
 $\phi_{0V}=0.5\pi$, $\phi_S=0.9\pi$, 
 $\om=0.2w$ and the time interval of one time period $[0,2\pi/\om]$ is divided into 
 50 intervals of equal size for calculating $U(T,0)$.}\label{fig:curr4}
\end{figure}

\begin{figure}
 \includegraphics[width=8cm]{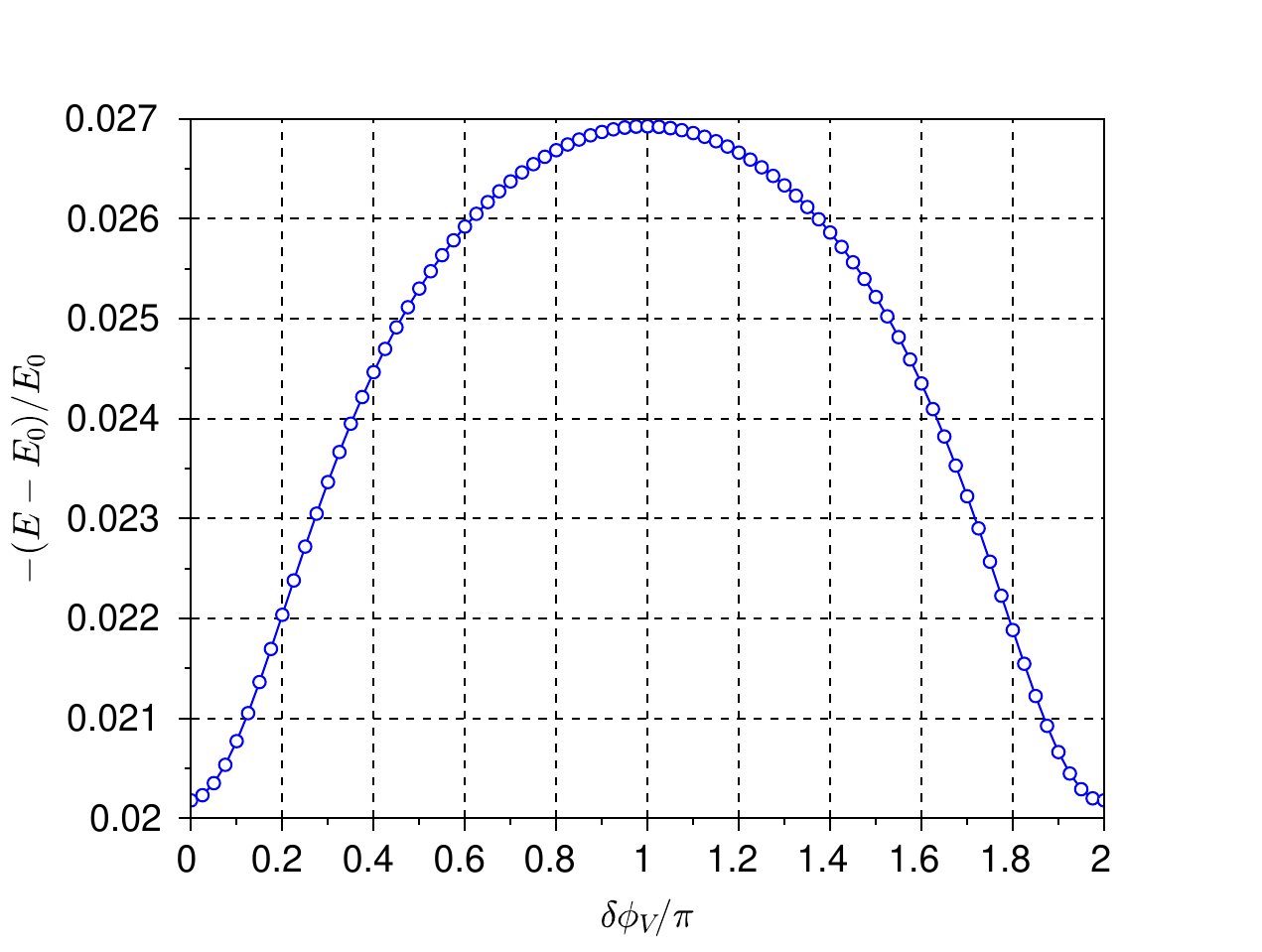}
 \caption{Time average of the energy of the driven system $E$ relative 
 to the ground state energy $E_0$ of the undriven system for the same parameters as in 
 Fig.~\ref{fig:curr4}}~\label{fig:dE4}
\end{figure}

Now, we switch on a superconducting phase difference $\phi_S=0.9\pi$ between the two 
superconductors which drives a Josephson current $I_J=- 0.08333 ew/\hbar$. Setting 
$\phi_{0V}=0.5\pi$, $\om=0.2w$ and keeping other parameters same as before, 
we calculate the  time averaged current as a function of the phase difference
between the oscillating potentials $\delta \phi_V$ and plot it in Fig.~\ref{fig:curr4}. 
We have chosen $\om$ to be close to  $2E_b=0.21823w$. The current $I_{av}$ which is 
the sum of the Josephson current and the pumped current deviates from the Josephson
current substantially for nonzero $\delta\phi_V$ while the deviation for zero 
$\delta\phi_V$ is minimal due to small $V_0$. Since the Josephson current carried
by different eigenstates of $H_0$ is substantially different for this choice of 
superconducting phase difference $\phi_S$, the  time averaged current $I_{av}$
can be expected to signify the excitations of the system. In Fig.~\ref{fig:dE4}, we plot
the excitation energy of the driven system relative to the ground state energy of the 
undriven system versus the difference in phases of the oscillating potentials
for the same set of parameters. 
We find that excitation energy shows the same dependence on the phase difference between
the oscillating potentials as the deviation of current $I_{av}$ of the driven system from 
the Josephson current $I_J$ of the undriven system. Thus, for a  value of 
superconducting phase difference $\phi_S$ so that $I_J\neq 0$, the deviation of 
current $I_{av}$ from  the Josephson current $I_J$ is a sign of excitations of the system. 
We find that such a correlation between the 
current $I_{av}$ and the excitations  of the driven system holds good for small amplitude 
of oscillating potential ($V_0<\De$).
\subsection{ Approaching the adiabatic limit}
\begin{figure}
 \includegraphics[width=4cm]{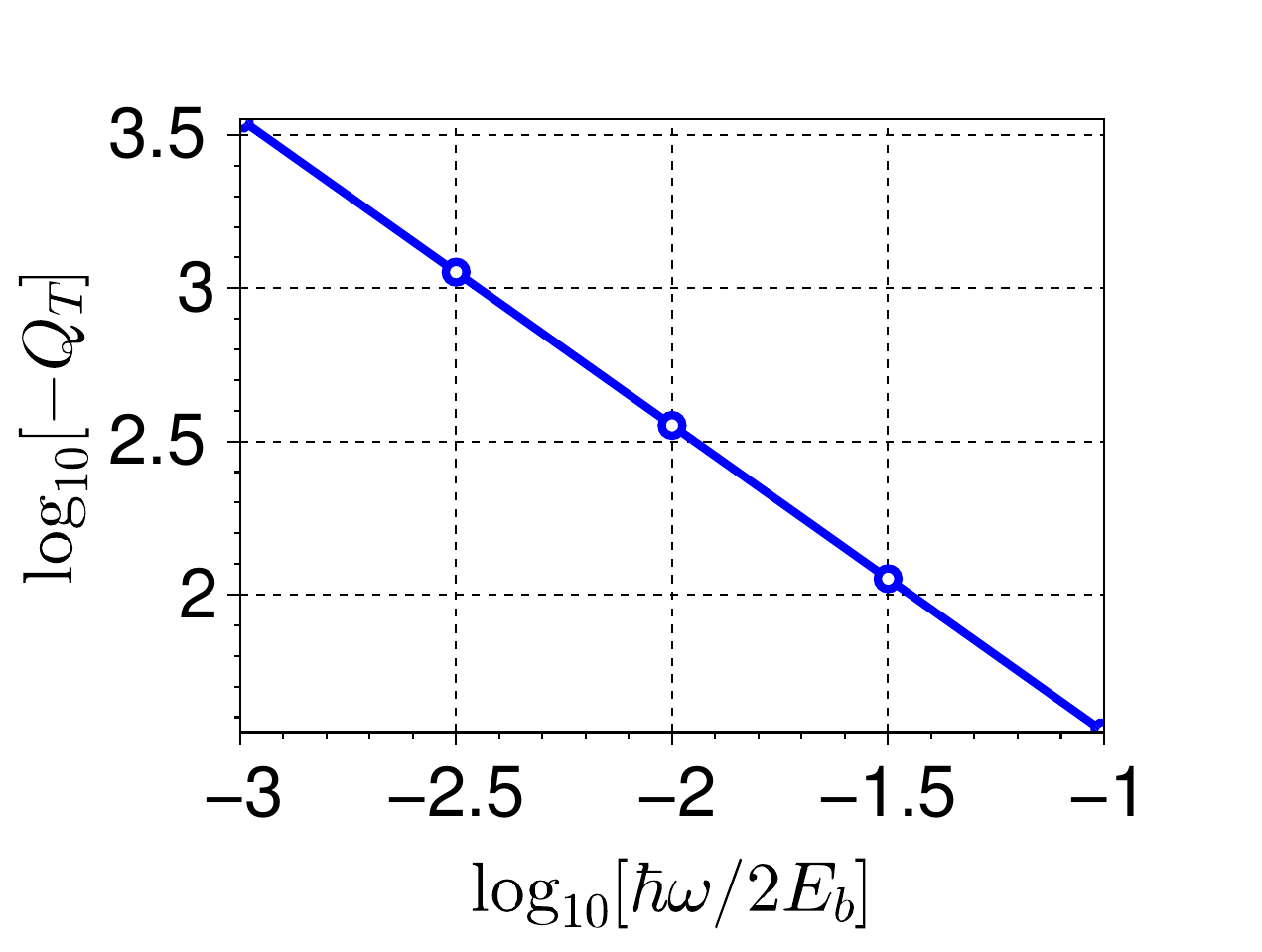}
 \includegraphics[width=4cm]{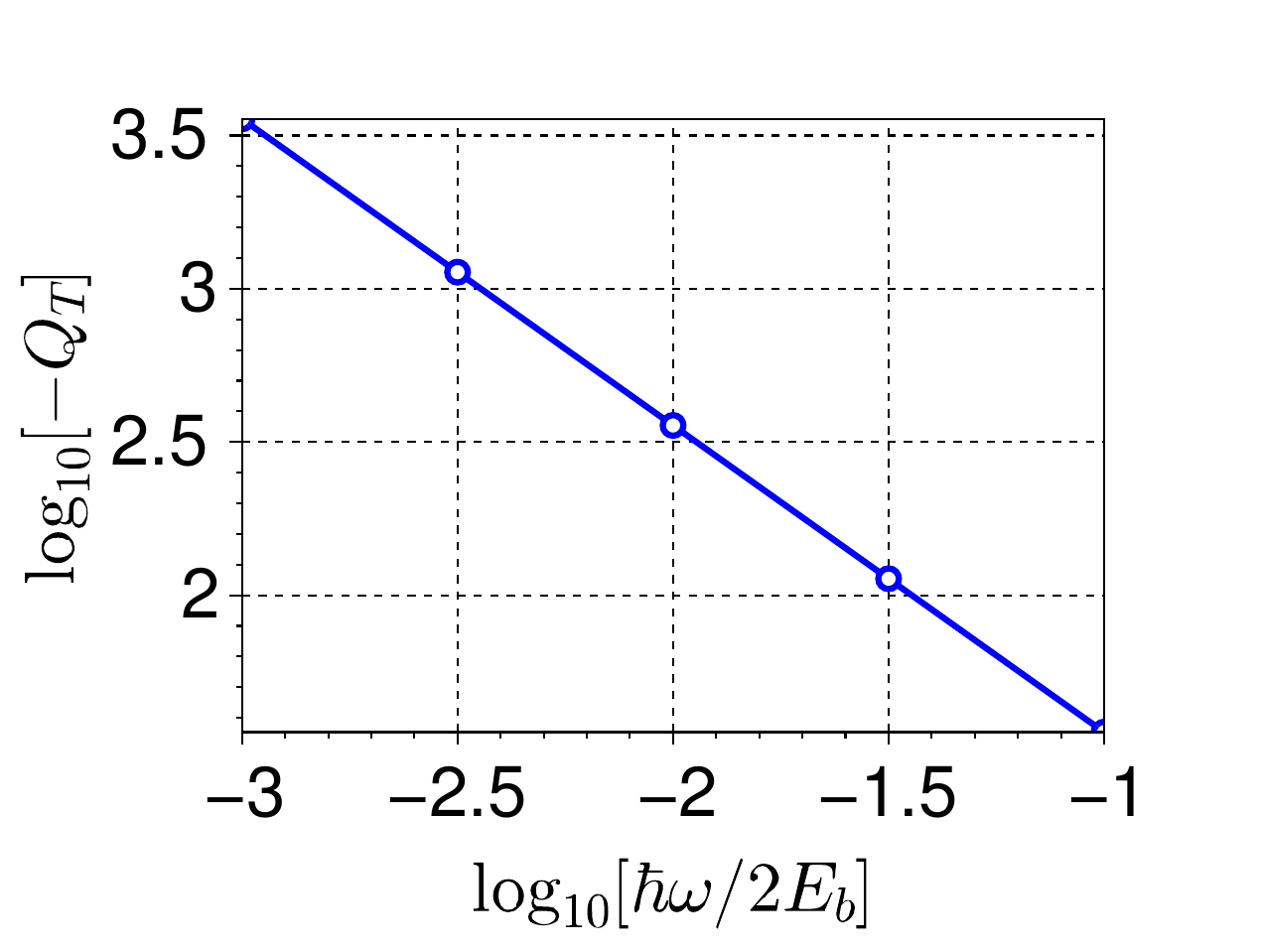}
 \caption{Logarithm of charge transferred per time period $T$ in units of $e$
 versus logarithm of pumping frequency $\omega$ for the choice of parameters: 
 $L_S=4$, $\mu=0.01w$, $\De=0.5w$, $w'=0.9$, $w''=0.3w$, $V_0=0.05w$, 
 $\phi_{0V}=0$,  and $\phi_S=0.8\pi$. Left panel: $\delta\phi_V=0.4\pi$, 
right panel: $\delta\phi_V=0$. A linear fit of  the form $y=mx+c$ to the 
curves with $(m,c,{\rm error})= (1.0004,0.5513,0.0004)$ for the left panel
and $(m,c,{\rm error})= (1.0003,0.5532,0.0003)$ for the right panel show 
that in both the cases, the charge transferred diverges in the limit $\om\to0$
as a powerlaw. }~\label{fig:adia}
\end{figure}

Now we turn our attention to the adiabatic limit. We study two site pumping since this 
can transfer nonzero pumped charge in the adiabatic limit when the superconducting phase
difference is zero. We take the adiabatic limit by taking the pumping frequency 
$\om\to0$. Since this implies that the 
time period $T\to\infty$, we maintain the size of the time slicing to be the same 
which means that the number of slices is proportional to the time period, 
given by $M=5[T\De/\hbar]$. Then, we calculate the  charge transferred in one time 
period (since the current $I_{av}$ may be zero in this limit) $Q_T = I_{av}T$. 
We choose the following parameters for two site pumping: 
$L_S=4$, $\mu=0.01w$, $\De=0.5w$, $w'=0.9$, $w''=0.3w$, $V_0=0.05w$ and 
 $\phi_{S}=0$.
 We numerically find that this gives a pumped charge that is zero in the limit 
$\om\to0$. Then, we choose $\phi_S=0.8\pi$ which drives a Josephson current of 
$I_J=-0.1406ew/\hbar$. Keeping other parameters the same, we drive the system at 
different frequencies and plot the charge transferred in one time period
$Q_T$ in Fig.~\ref{fig:adia} for $\delta\phi_V=0.4\pi$ (in left panel) and 
for $\delta\phi_V=0$ (in right panel) choosing $\phi_{0V}=0$. 
We find that the pumped charge diverges when $\om\to0$ as a powerlaw
given by $Q_T=Q_0 (\hbar\om/2E_b)^{-m}$. This form of dependence is not 
surprising given that $m\simeq 1$. If only the Josephson current flows across, 
$Q_T=I_JT=(\pi\hbar I_J/E_b)\cdot(\hbar\om/2E_b)^{-1}$. Hence, deviation of 
$m$ from unity and $Q_0$ from $\pi\hbar I_J/E_b$ indicates that the charge 
transferred is not purely by the Josephson current of the undriven system. 
Here, we find $\pi\hbar I_J/E_b=-3.6267e$. From the curve fitting in 
Fig~\ref{fig:adia}, $Q_0=-3.5588e$ for $\delta\phi_V=0.4\pi$ and 
$Q_0=-3.5744e$ for $\delta\phi_V=0$. Interestingly, even pumping at two sites with a 
difference in phases of the oscillating potentials $\delta\phi_V=0$ marks a 
deviation of $Q_0$ from $\pi\hbar I_J/E_b$. Such a deviation grows in magnitude 
with increasing $V_0$.  This is possibly due to the
excitations in the system that get carried all the way down to the adiabatic 
limit. Further, subtracting out the charge transferred solely by the 
Josephson current still gives a powerlaw dependence of similar form for the 
remnant charge. Hence, the charge transferred purely due to pumping in this case
(for nonzero $I_J$) cannot be neglected in the adiabatic limit. 

\subsection{ Dependence on $L_S$}
For small $L_S$, the boundary effects become important. They even change the 
subgap energies $\pm E_b$. For large $L_S$, the subgap energies do not change with $L_S$. For the 
choice of parameters here, $L_S=4$ is the borderline above which subgap energies do not change with 
$L_S$. Coming to the results, the results on current $I_{av}$ do not change for $L_S\ge4$ when 
$V_0<\Delta$ and $\hbar\omega<\Delta$. If either $V_0>\Delta$ or $\hbar\omega>\Delta$ the results
depend on $L_S$ since the eigenenergies of the time-independent Hamiltonian outside the gap change
with $L_S$ and they participate in the transport.

\section{Discussion and Conclusion}~\label{sec-con}
A closely related phenomenon is inverse ac Josephson effect~\cite{Langenberg66} in which 
a microwave electromagnetic radiation shined on the Josephson junction results in 
a time-varying voltage whose dc component is quantized. However, a  sinusoidally 
varying current is induced in this effect which results in variation of Josephson
phase difference with time in contrast to our work where the Josephson phase difference 
remains fixed. The time dependent potentials in our work create an excited Floquet state
for the hybrid device but not a potential difference between one superconductor and the other. 

It is known that periodically driven interacting systems approach a periodic steady 
state~\cite{laza14prl,russomanno12}. So, due to weak interactions in the realistic system,
the system approaches a periodic steady state described by the Floquet states after a long
time. The current averaged over one time period in such a state will be $I_{av}$. And the
excitation energy averaged over one time period will be $E$ given by eq.~\eqref{eq:ET}. 
Hence, $I_{av}$ can be measured in a realistic system. Our prediction that deviation of
time averaged current $I_{av}$ of the driven system from the Josephson current $I_J$
of the undriven system  is a signature of excitations of the driven system (when $I_J\neq
0$) can be applied to realistic systems.   Identification of a correlation between 
physical quantities of interest in Floquet systems to a measurable physical  quantity
such as current is an important step that can improve our understanding of periodically
driven systems. 

The total current in the Josephson junction with an applied oscillating potential can be 
seen as a sum of the Josephson current and a pumped current. The pumped current 
can be nonzero when spatial inversion symmetry is broken (by a nonzero superconducting
phase difference) and a time dependent potential is applied to one site. 
We see from  Fig.~\ref{fig:curr1} and Fig.~\ref{fig:curr2}
that this is indeed the case. Moreover,  at some values of the superconducting phase 
difference the pumped current is not only in opposite direction to the Josephson current 
but the former is greater in magnitude than the latter. This is analogous to the 
current reversal by shining microwave radiation on Josephson junction proposed by 
Gorelik~et.al.~\cite{gorelik95}. Further, we have shown that 
by applying oscillating potentials at two sites differing by a nonzero phase, 
charge can be pumped between two superconductors maintained at same superconducting
phase. In the adiabatic limit, we find that charge gets transferred in one time period 
when a Josephson current flows from one superconductor to another in the undriven 
system. This charge diverges as a powerlaw with the pumping frequency in the 
adiabatic limit. The charge transferred so is predominantly carried by the Josephson
current of the undriven system but with a correction. However, the charge transferred
in one time period purely due to pumping cannot be said to be zero. 

We have described a way of calculating pumped current exactly for a nonadiabatically 
driven Josephson junction. To test our predictions experimentally, we propose to
connect superconducting leads to the electrostatically defined semiconductor quantum 
dots where the gate voltages can be used to apply time dependent potentials.
Charge pumping has been achieved in quantum dots synthesized  from GaAs-AlGaAs 
heterostructures (except that the reservoirs are normal metal leads)~\cite{switkes99}.
Further, superconductivity can be induced in GaAs heterostructures~\cite{Wan2015}. Hence, 
gate tunable quantum dots can be connected to superconducting leads. 
Another direction to experimental realization comes from the
fact that carbon nanotubes can be employed as quantum dots and certain superconductors
can be connected to carbon nanotubes~\cite{Cleuziou2006}. Apart from these, there are 
many more platforms where quantum dots have been coupled to 
superconductors~\cite{DeFranceschi2010}. In these systems, the quantum dots can be gate 
tuned. AC voltage source need to be connected to the gate electrodes to induce time dependent
potentials in  the normal metal region. The  task in engineering the system we studied is 
to maintain a tunable Josephson phase difference between the two finite superconductors 
connected on either sides of the quantum dot.
We envisage that such periodically driven Josephson junctions will be realized
in near future and our predictions can be tested. 

\acknowledgements
A.S. thanks Diptiman Sen and Dhavala Suri for fruitful discussions. 
A.S. thanks DST-INSPIRE Faculty Award (Faculty Reg. No.~:~IFA17-PH190)
for financial support. 
\bibliography{refpump}

\end{document}